\documentclass[twocolumn]{aastex631}

\usepackage[utf8]{inputenc} 
\usepackage[T1]{fontenc}    
\usepackage{hyperref}       
\usepackage{url}            
\usepackage{booktabs}       
\usepackage{amsfonts}       
\usepackage{nicefrac}       
\usepackage{amsmath}
\usepackage{amssymb}
\usepackage{multirow}
\usepackage{float}
\usepackage{graphicx}
\usepackage{cleveref}
\usepackage{physics}
\usepackage{newtxtext,newtxmath}

\newcommand{\tilxi}{\Tilde{\xi}} 
\newcommand{\be}{\begin{equation}} 
\newcommand{\ee}{\end{equation}} 
\newcommand{\bary}{\begin{eqnarray}} 
\newcommand{\eary}{\end{eqnarray}} 
\Crefname{equation}{Eq.}{Eqs.} 

\begin{document}

\title[GRB 190829A]{Exploring the Early Afterglow Polarization of GRB 190829A}

\author[0000-0001-5785-8305]{A. C. Caligula Do E. S. Pedreira}
\affiliation{Instituto de Astronom\'ia, Universidad Nacional Aut\'onoma de M\'exico, Circuito Exterior, C.U., A. Postal 70-264, 04510, CDMX, Mexico}
\author[0000-0002-0173-6453]{N. Fraija}
\affiliation{Instituto de Astronom\'ia, Universidad Nacional Aut\'onoma de M\'exico, Circuito Exterior, C.U., A. Postal 70-264, 04510, CDMX, Mexico}
\author[0000-0001-6849-1270]{S. Dichiara}
\affiliation{Department of Astronomy and Astrophysics, The Pennsylvania State University, 525 Davey Lab, University Park, PA 16802, USA}
\author[0000-0002-2149-9846]{P. Veres}
\affiliation{Center for Space Plasma and Aeronomic Research (CSPAR), University of Alabama in Huntsville, Huntsville, AL 35899, USA}
\author[0000-0003-4442-8546]{M.G. Dainotti}
\affiliation{Division of Science, National Astronomical Observatory of Japan, 2-21-1 Osawa, Mitaka, Tokyo 181-8588, Japan}
\affiliation{The Graduate University for Advanced Studies (SOKENDAI),
2-21-1 Osawa, Mitaka, Tokyo 181-8588, Japan}
\affiliation{Space Science Institute, 4750 Walnut Street, Boulder, CO
80301, USA}
\affiliation{SLAC National Accelerator Laboratory, 2575 Sand Hill Road, Menlo Park, CA 94025, USA}
\author[0000-0001-5193-3693]{A. Galvan-Gamez} 
\affiliation{Instituto de Astronom\'ia, Universidad Nacional Aut\'onoma de M\'exico, Circuito Exterior, C.U., A. Postal 70-264, 04510, CDMX, Mexico}
\author[0000-0002-0216-3415]{R.~L.~Becerra}
\affiliation{Instituto de Ciencias Nucleares, Universidad Nacional Aut\'onoma de M\'exico, Apartado Postal 70-264, 04510 M\'exico, CDMX, Mexico}
\author[0000-0002-2516-5739]{B. Betancourt Kamenetskaia}
\affiliation{TUM Physics Department, Technical University of Munich, James-Franck-Str, 85748 Garching, Germany}
\affiliation{Max-Planck-Institut für Physik (Werner-Heisenberg-Institut), Föhringer Ring 6, 80805 Munich, Germany}

\shortauthors{Caligula et al.}

\begin{abstract}
The GRB 190829A has been widely studied due to its nature and the high energy emission presented.
Due to the detection of a very-high-energy component by the High Energy Stereoscopic System and the event's atypically middling luminosity, it has been categorized in a select, limited group of bursts bordering classic GRBs and nearby sub-energetic events. Given the range of models utilized to adequately characterize the afterglow of this burst, it has proven challenging to identify the most probable explanation.
Nevertheless, the detection of polarization data provided by the MASTER collaboration has added a new aspect to GRB 190829A that permits us to attempt to explore this degeneracy. In this paper, we present a polarization model coupled with a synchrotron forward-shock model – a component in all models used to describe GRB 190829A's afterglow -- in order to fit the polarization's temporal evolution with the existing upper limits ($\Pi < 6\%$).
We find that the polarization generated from an on-axis emission is favored for strongly anisotropic magnetic field ratios, while an off-axis scenario cannot be fully ruled out when a more isotropic framework is taken into account. 
\end{abstract}


\section{Introduction}\label{sec1}
GRBs (gamma-ray bursts) are among the brightest phenomena in the universe. They are originated when massive stars die \citep{1993ApJ...405..273W,1998ApJ...494L..45P, 2006ARA&A..44..507W,Cano2017} or two compact objects, such as neutron stars \citep[NSs;][]{1992ApJ...392L...9D, 1992Natur.357..472U, 1994MNRAS.270..480T, 1989Natur.340..126E, 2011MNRAS.413.2031M} and a NS - black hole \citep[BH,][]{1992ApJ...395L..83N}, merge. Long GRBs (lGRBs) and short GRBs (sGRBs), usually associated with the dying of massive stars and a merging of compact objects, are commonly classified based on their duration:\footnote{For a debate of controversial situations, see \cite{kann2011,becerra2019b}.} $T_{90}\leq 2\mathrm{\,s}$ or $T_{90} \ge 2\mathrm{\,s}$,\footnote{$T_{90}$ is the time over which a GRB releases from $5\%$ to $95\%$ of the total measured counts.} respectively \citep{mazets1981catalog, kouveliotou1993identification}. GRBs are studied according to their phenomenology detected during the early and late temporal phases. From hard X-rays to $\ge100$ MeV $gamma$-rays, the early and main episode known as the ``prompt emission" is detected and explained by interactions with internal shells of material launched by a central engine at varying velocities and therefore, at different Lorentz factors \citep{1994ApJ...430L..93R, 1994ApJ...427..708P}, the photospheric emission \citep{2007ApJ...666.1012T,2011ApJ...732...26M,2013ApJ...765..103L} or discharges from a Poynting-flux dominated outflow \citep{2008A&A...480..305G, 2011ApJ...726...90Z,2015MNRAS.453.1820K,2016MNRAS.459.3635B}. The long-lasting multi-wavelength emission observed in $\gamma$-rays, X-rays, optical, and radio is known as ``afterglow" \citep[e.g.,][]{1997Natur.387..783C, 1997Natur.386..686V,1998A&A...331L..41P,1998ApJ...497L..17S,2002ApJ...568..820G, Gehrels2009ARA&A,Wang2015}. It is often modelled with synchrotron radiation occurring when the relativistic outflow is decelerated by an external medium, and a significant portion of its energy is transferred to it \citep[e.g., see][]{1998ApJ...497L..17S}.

The synchrotron radiation produced at the afterglow is typically explored in a forward-shock (FS) scenario, and this mechanism is contingent on the existence of magnetic fields in the emission zone. The origin of magnetic fields in GRBs is still a topic of discussion. Magnetic fields can be advected from the burst's source \citep{2003astro.ph.12347L, 2011ApJ...726...75S}, be generated from compression of an existing interstellar medium (ISM) magnetic field \citep{1980MNRAS.193..439L,2021MNRAS.507.5340T}, from shock-generated plasma instabilities \citep{PhysRevLett.2.83, Medvedev}, and even by magnetic reconnection in Poyinting-flux dominated outflows \citep{2011ApJ...726...90Z}. Probing these fields, and their originating sources, alongside other properties of GRBs, is a challenging task that requires multiple directions of theoretical considerations and observational data. One way is exploring the polarization characteristics of an observed emission. As the polarization relies on the configuration of the magnetic field, it analysis allow us to look into their configurations and, therefore, their origins. 

Several authors have implemented polarization models in order to acquire source-related information \citep[e.g., see][]{2003ApJ...594L..83G, Lyutikov, Nakar, 2004MNRAS.354...86R, Gill-1,  2020ApJ...892..131S, 2021MNRAS.507.5340T, 2022MNRAS.tmp.2180S}. Unfortunately, despite the number of GRB observations, collecting polarization data remains being a challenge. Nevertheless, afterglow polarization measurements have been obtained for several GRBs at this point: at the early afterglow phase, the observations of GRB 090102 ($\Pi = 10.2\pm1.3\%$, \cite{Steele}), GRB 120308A ($\Pi = 28\pm4\%$, \citet{Mundell}) and the upper limits of GRB 190829A ($\Pi < 6\%$, \cite{2022MNRAS.512.2337D}) were obtained; examples at the late afterglow include GRB 191221B \citep[$\Pi=1.2\%$;][]{Buckley}, GRB 190114C \citep[$\Pi=0.8\pm0.13\%$;][]{Laskar_2019} on the radio band, and the upper limits of GRB 991216 \citep[with $\Pi < 7\%$;][]{2005ApJ...625..263G} and GRB 170817A \citep[with $\Pi < 12\%$, on the 2.8 GHz radio band][]{2018ApJ...861L..10C}. Furthermore, several global collaborations like the POLAR project \citep{POLAR}, the Multicolour OPTimised Optical Polarimeter \citep[MOPTOP;][]{2020MNRAS.494.4676S}, and the MASTER project \citep{2019ARep...63..293L}, which are underway, or in preparation phases, could provide future observations and the much-needed polarization data of different GRB epochs.

In this work, we calculate the temporal evolution of the expected polarization for GRB 190829A in a synchrotron FS scenario, proven to allow for analysis of GRB 190829A's afterglow in terms of the physical parameters of the system. We use the available polarimetric upper limits from \cite{2022MNRAS.512.2337D} to test the parameters and models required to explain the multi-wavelength afterglow observations of GRB 190829A. The paper is structured as follows; In Section \ref{sec2}, we briefly introduce the polarization model used in this paper in the synchrotron FS framework during the off-axis phase. In Section \ref{sec3}, we apply as particular case our polarization model to GRB 190829A using the best-fit parameters obtained in different scenarios. In Section \ref{sec4}, we discuss the results. Finally, in Section \ref{sec7}, we offer a brief summary and our concluding remarks.

\section{Linear Polarization from off-axis Synchrotron afterglow}\label{sec2}

\subsection{Linear Polarization model}

Polarization is commonly attributed to synchrotron radiation behind shock waves. This makes it dependent on the arrangement of the magnetic field and the geometry of the shock, as they determine the degree of polarization ($\Pi$) in each position as well the integrated value throughout the entire image \citep{Gill-1}.
The Stokes parameters ($I$, $Q$, $U$, and $V$) set the method for calculating polarization, and only linear polarization is generally taken into account \citep[see][for an analysis of circular polarization in GRBs]{2016MNRAS.455.1594N}.
Hereinafter, quantities in the observer and comoving frames will be referred to as unprimed and primed, respectively.
The linear polarization Stokes parameters are written as: 

\begin{eqnarray}
   &V = 0, \hspace{2cm} &\theta_p = \frac{1}{2}\arctan{\frac{U}{Q}},\,\cr
    &\frac{U}{I} = \Pi'\sin{2\theta_p}, \hspace{1cm}
    &\frac{Q}{I} = \Pi'\cos{2\theta_p}.
\end{eqnarray}

The sum over the flux returns the measured Stokes parameters \citep{Granot-P2}, so:
 
\begin{eqnarray}
       &\frac{U}{I} = \frac{\int \mathrm{d}F_\nu\Pi'\sin{2\theta_p}}{\int \mathrm{d}F_\nu},\hspace{1cm}
    \frac{Q}{I} =  \frac{\int \mathrm{d}F_\nu\Pi'\cos{2\theta_p}}{\int \mathrm{d}F_\nu},  \\
   &\Pi = \frac{\sqrt{Q^2+U^2}}{I}.
\end{eqnarray}

Considering $\mathrm{d}F_\nu \propto   \delta_D^3L'_{\nu'}\mathrm{d}\Omega$ -- where $L'_{\nu'}$ is the spectral luminosity, $\delta_D$ is the Doppler factor and $\mathrm{d}\Omega$ is the element of solid angle of the fluid element in relation to the source -- the factors regarding the geometry of the magnetic field and outflow can be introduced by \citep{ribicky} are:
\begin{eqnarray}
    L'_{\nu'} \propto (\nu')^{-\alpha} (\sin{\chi'})^\epsilon r^m \propto (\nu')^{-\alpha} (1-\hat{n}' \cdot \hat{B}')^{\epsilon/2} r^m.
\end{eqnarray}
The term $\epsilon = 1+\alpha$, with $\alpha$ the spectral index, which can be obtained with the electron power-law index $p$; i.e., $\alpha = (p-1)/2$ \citep{Granot-P2}. Throughout the text, we also assume the index $m=0$, i.e., a radially constant emissivity.

The angle between the local magnetic field and the direction of motion of the particle, $\chi$, is also the pitch angle due to the highly beamed effect of the synchrotron emission. The geometrical considerations of polarization can then be taken by averaging this factor over the local probability distribution of the magnetic field \citep{Gill-1},

\begin{eqnarray}
    \Lambda = \expval{(1-\hat{n}' \cdot \hat{B}')^{\epsilon/2}}.
\end{eqnarray}

Using the Lorentz transformation of the unit vector $\hat{n}$ or a prescription of $\hat{B}$, it is possible to obtain $\Lambda$ in terms of different magnetic field configurations and minimize the number of parameters required to obtain the polarization degree \citep{Lyutikov, Granot-P2, 2005ApJ...625..263G, Gill-1}:

\begin{eqnarray}
    \Lambda_{\perp} &\approx& \expval{\left[\left(\frac{1-\Tilde{\xi}}{1+\Tilde{\xi}}\right)\cos^2{\varphi_B} + \sin^2{\varphi_B}\right]^{\epsilon/2}}_{\varphi_B}\label{eq:pd_perp},\\
    \Lambda_{\parallel} &\approx& \left[\frac{\sqrt{4\Tilde{\xi}}}{1+\Tilde{\xi}}\right]^\epsilon \label{eq:pd_par},\\
    \Lambda_{tor} &\approx& \left[\left(\frac{1-\Tilde{\xi}}{1+\Tilde{\xi}}\right) + \frac{{4\Tilde{\xi}}}{(1+\Tilde{\xi})^2} \frac{(a+ \cos{\Tilde{\varphi}})^2}{1+a^2 + 2a\cos{\Tilde{\varphi}}}\right]^{\epsilon/2}\label{eq:pd_tor},
\end{eqnarray}

where $\varphi_B$ is the azimuthal angle of the magnetic field measured from a reference point; $\tilde{\xi} \equiv (\Gamma\tilde{\theta})^2$, taking in consideration the approximations of $\Tilde{\mu} = \cos{\tilde{\theta}} \approx 1 - \tilde{\theta}^2/2$ and $\beta \approx 1 - 1/2\Gamma^2 $, which leads to $ \delta_D \approx \frac{2\Gamma}{1+\Tilde{\xi}}$; and $a\equiv \tilde{\theta}/\theta_{obs}$, where $\tilde{\theta}$ is the polar angle measured from the Line of Sight (LOS), $\Gamma$ is the bulk Lorentz factor, and $\beta$, is the velocity of the material in terms of the speed of light.

The uncertainty regarding the present magnetic field configuration in the region of emission requires the exploration of multiple configurations, as well as consider the relevance of magnetic field geometry regarding the polarization degree evolution. For the early afterglow, three of the most suitable configurations are: a random perpendicular configuration -- where the anisotropy factor $b \equiv \frac{2\expval{B_\parallel^2}}{\expval{B_\perp^2}}= 0$ -- confined to the shock plane; an ordered configuration parallel to the velocity vector, where $b\rightarrow\infty$; and an ordered toroidal magnetic field configuration arising from an axisymmetric field configuration with a poloidal $(B_{\rm p} \propto r^{-2})$ and toroidal component $(B_{\rm \phi} \propto r^{-1})$. More complex configurations with multi-component, where the anisotropy is more generalized, magnetic fields have been explored \citep{2018ApJ...861L..10C, 2020ApJ...892..131S, 2020MNRAS.491.5815G, 2021MNRAS.507.5340T}, as it is warranted and needed, however, for the purposes of this paper we limit ourselves to the three following cases.

\paragraph{Random magnetic field -- Perpendicular ($B_\perp,\,b=0$).}
In this scenario, the symmetry of the random magnetic field configuration, perpendicular to the shock plane, causes the disappearance of the polarization over the image if the beaming cone is wholly contained within the jet aperture or if it is seen along the axis ($\theta_{\rm obs} = 0$). To break the symmetry, the jet must be viewed close to its edge ($q\equiv \frac{\theta_{\rm obs}}{\theta_{\rm j}} \gtrsim 1+\xi_j^{-1/2}$), where missing emission (from $\theta > \theta_{\rm j}$) results only in partial cancellation \citep{Waxman}.
The equation necessary to calculate this polarization is explicitly laid out as Eq. 5 in \cite{Granot-P2}.


\paragraph{Ordered magnetic field -- Parallel ($B_\parallel, \,b\rightarrow\infty$).}
For the ordered magnetic field, a configuration parallel to the velocity vector, the same symmetry observations hold true and the calculation follows \cite{Granot-P2, Gill-1}, with $\Lambda(\tilxi) = \Lambda_\parallel$ from Eq. ~\ref{eq:pd_par}.

\paragraph{Ordered magnetic field -- Toroidal ($B_{\rm tor}$).}
For the ordered magnetic field in a toroidal configuration, the same symmetry concerns are maintained. The calculation follows \cite{2005ApJ...625..263G, Gill-1}, with $\Lambda(\tilxi) = \Lambda_{\rm tor}$ from Eq. ~\ref{eq:pd_tor}.


By substituting the following integration limits:

\begin{eqnarray}\label{conections}
    &\cos{\psi(\tilxi)} = \frac{(1-q)^2 \xi_j - \tilxi}{2q\sqrt{\xi_j\tilxi }}, \\
    &\qquad \xi_j = (\Gamma\theta_{\rm j})^2, \qquad \xi_\pm = (1\pm q)^2\xi_j,
\end{eqnarray}

with an appropriate prescription of the bulk Lorentz factor $\Gamma(t)$, the evolution of the opening angle of the jet $\theta_{\rm j}(t)$, and the parameters required to describe these expressions, we can obtain the temporal evolution of polarization.


\subsection{Synchrotron off-axis afterglow scenario: An homogeneous case}\label{subsec32}


We assume an adiabatic evolution of the forward shock in a homogeneous medium $n$ with an isotropic equivalent-kinetic energy  $E=\frac{\Omega}{3}r^{3}m_p c^2 n  \Gamma^2$ \citep[Blandford-McKee solution;][]{1976PhFl...19.1130B} and a radial distance $r=c\beta t/[(1+z)(1-\beta\mu)]$. Then, the evolution of the bulk Lorentz factor is given by: 

\bary
\label{Gamma_dec_off}
\Gamma = \left(\frac{3}{4\pi\,m_p c^{5}}\right)^{\frac12} \,(1+z)^{\frac{3}{2}}  (1-\beta\cos\Delta\theta)^{\frac{3}{2}}\,n^{-\frac{1}{2}}\,E^{\frac{1}{2}}t^{-\frac{3}{2}} \,,
\eary

with $\beta=\sqrt{\Gamma^2-1}/\Gamma$, $\Delta\theta= \theta_{\rm obs} - \theta_{\rm j}$, $\Omega$ is the solid angle, $m_p$ is the proton mass and $c$ is the speed of light. In forward-shock models, accelerated electrons are described by taking into account their Lorentz factors ($\gamma_e$) and the electron power index $p$. This leads to a distribution of the form $N(\gamma_e)\,d\gamma_e \propto \gamma_e^{-p}\,d\gamma_e$ for $\gamma_m\leq \gamma_{\rm e}$, where $\gamma_m=m_{\rm p}/m_{\rm e}g(p)\varepsilon_{\rm e}(\Gamma-1)\zeta^{-1}_e$ is the minimum electron Lorentz factor with $m_{\rm e}$ the electron mass, $\varepsilon_{\rm e}$ the fraction of energy given to accelerate electrons, $\zeta_{e}$ the fraction of electrons that were accelerated by the shock front \citep{2006MNRAS.369..197F} and $g(p)=\frac{p-2}{p-1}$. The comoving magnetic field strength in the blast wave can be expressed as $B'^2/(8\pi)=\varepsilon_Be$, where knowledge of the energy density $e=[(\hat\gamma\Gamma +1)/(\hat\gamma - 1)](\Gamma -1) n m_pc^2$, adiabatic index $\hat\gamma$ \citep{1999MNRAS.309..513H}, fraction of energy provided to the magnetic field ($\varepsilon_B$) is necessary with $n$ the constant-density medium of the circumburst environment. The cooling electron Lorentz factor is written as $\gamma_{\rm c}=(6\pi m_e c/\sigma_T)(1+Y)^{-1}\Gamma^{-1}B'^{-2}t^{-1}$, where $\sigma_T$ is the Thomson cross-section and $Y$ is the Compton parameter \citep{2001ApJ...548..787S, 2010ApJ...712.1232W}. The synchrotron spectral breaks can now be expressed in terms of previously defined quantities as $\nu'_{\rm i}=q_e/(2\pi m_ec)\gamma^{2}_{\rm i}B'$, where the sub-index ${\rm i=m}$ and ${\rm c}$ will stand for the characteristic or cooling break, respectively. The constant $q_e$ represents the elementary charge. The synchrotron radiation power per electron in the comoving frame is given by $P'_{\nu'_m}\simeq \sqrt{3}q_e^3/(m_ec^2)B'$ \citep[e.g., see][]{1998ApJ...497L..17S, 2015ApJ...804..105F}.  Considering the total number of emitting electrons $N_e=\frac{\Omega n r^3}{3}$ and also taking into account the transformation laws for the solid angle ($\Omega= \Omega'/\delta^2_D$), the radiation power ($P_{\nu_m}=\delta_D/(1+z) P'_{\nu'_m}$) and the spectral breaks ($\nu_{\rm i}=\delta_D/(1+z)\nu'_{\rm i}$), the maximum flux given by synchrotron radiation is:

\bary
F_{\rm \nu, max}=\frac{(1+z)^2\delta^3_D}{4\pi d_z^2}N_eP'_{\nu'_m}\,,
\eary 

where {\small $d_{\rm z}=(1+z)\frac{c}{H_0}\int^z_0\,\frac{d\tilde{z}}{\sqrt{\Omega_{\rm M}(1+\tilde{z})^3+\Omega_\Lambda}}$} \citep{1972gcpa.book.....W} is the luminosity distance. For the cosmological constants, we assume a spatially flat universe $\Lambda$CDM model with $H_0=69.6\,{\rm km\,s^{-1}\,Mpc^{-1}}$, $\Omega_{\rm M}=0.286$ and $\Omega_\Lambda=0.714$ \citep{2016A&A...594A..13P}. 


\section{Application: GRB 190829A} \label{sec3}

\subsection{The Prompt Episode} \label{subsec30}
GRB 190829A was triggered and located by the {\itshape Fermi}/Gamma-ray Monitor (GBM) and {\itshape Swift}/Burst Alert Telescope (BAT) instruments at $T = $ 2019 August 29 19:55:53.13 UTC and 19:56:44.60 UTC, respectively \citep{2019GCN.25551....1F,2019GCN.25552....1D}. {\itshape Swift}/BAT localized GRB 190829A with coordinates RA, DEC(J2000)=44.540, -8.968 with a 90\% error radius of 3 arcmin. The {\itshape Fermi}/GBM light curve exhibited an initial pulse followed by a brighter peak released total isotropic-equivalent energies of $E_{\rm \gamma, iso}=(9.151\pm 0.504)\times 10^{49}$ erg and $(2.967\pm 0.032)\times 10^{50}$ erg, respectively \citep{2021ApJ...918...12F}. The {\itshape Fermi}/GBM and {\itshape Swift}/BAT observations are presented and interpreted in several scenarios \citep[e.g., see][]{2021ApJ...918...12F, 2020ApJ...898...42C, 2021ApJ...917...95Z, 2021MNRAS.504.5647S, 2022ApJ...931L..19S}.\\ 

\subsection{The Afterglow Phase} \label{subsec31}

Long term follow-up observations of this burst were performed by X-ray Telescope (XRT) and UltraViolet Optical Telescope (UVOT) on board the Neil Gehrels Swift Observatory, and MASTER Telescopes \citep{dichiara2021}. The {\itshape Swift}/XRT instrument started detecting this burst at 19:58:21.9 UTC, 97.3~s after the BAT trigger. The observations with {\itshape Swift}/UVOT at 106~s after the BAT trigger. This instrument detected in V, B, white, U, UVW1, UVW2 and UVM2 filters exhibiting a bright flare peaking at $\sim 2\times 10^3\,{\rm s}$. The MASTER network observed GRB 190829A at $T+1239$~s after the burst trigger time. GRB 190829A was observed with the Clear (C) and the polarization filters (P). The optical fluxes from the P filter corresponds to superposition of B and R standard Johnson filters (0.2 B + 0.8 R) \citep{2022MNRAS.512.2337D}.

The multi-wavelength afterglow observations of GRB 190829A were modelled by several collaborations \citep{2020MNRAS.496.3326R,2021MNRAS.504.5647S,2022arXiv220813987S, 2021ApJ...918...12F, 2022ApJ...931L..19S, 2022MNRAS.512.2337D, 2021ApJ...917...95Z, 2021ApJ...920...55Z}. These observations have shown to be consistent with a FS plus reverse shock (RS) scenario with Synchrotron Self-Compton (SSC) emission \citep{2020MNRAS.496.3326R,2022ApJ...931L..19S, 2022MNRAS.512.2337D}. On the other hand, \cite{2021MNRAS.504.5647S,2022arXiv220813987S} applied a dual component model -- with a narrow off-axis jet describing the VHE component and early afterglow; while a wide-angle, slower, jet seen closer to its edge describes the late time afterglow observations, to model the complex afterglow of GRB 190829A. \cite{2021ApJ...918...12F} presented a synchrotron forward shock model originating from a spin-down millisecond magnetar source as a strong candidate for this burst. \cite{2021ApJ...917...95Z} defined a set of archetypical rules and applied them to a synchotron FS + SSC model to describe GRB 190829A's observations; while \cite{2021ApJ...920...55Z} shown the afterglow to be compatible with a synchrotron FS + inverse Compton (iC) case, assuming both SSC and external inverse Compton (eiC) components.

This kind of degeneracy between models is expected in GRB afterglows \citep{2004MNRAS.354...86R, 2009MNRAS.400L..75K, 2010MNRAS.409..226K, 2017ApJ...848...15F}, therefore, the approach of a scenario requires a profound analysis of each possibility. GRB 190829A had observed $\abs{Q}$ values, and consequently, upper limits imposed upon its early afterglow \citep{2022MNRAS.512.2337D} and fitting this data can be useful in an attempt to break the models' degeneracy. In the following subsections we present our semi-analytical model, an off-axis top-hat jet with synchrotron FS dynamics, similar to the ones presented by the authors, mentioned before, to describe the FS, which we use to calculate the polarization evolution of this burst, using the parameters obtained by the aforementioned works.

\section{Results and Discussion}\label{sec4}

\begin{figure}
{\includegraphics[width=\textwidth]{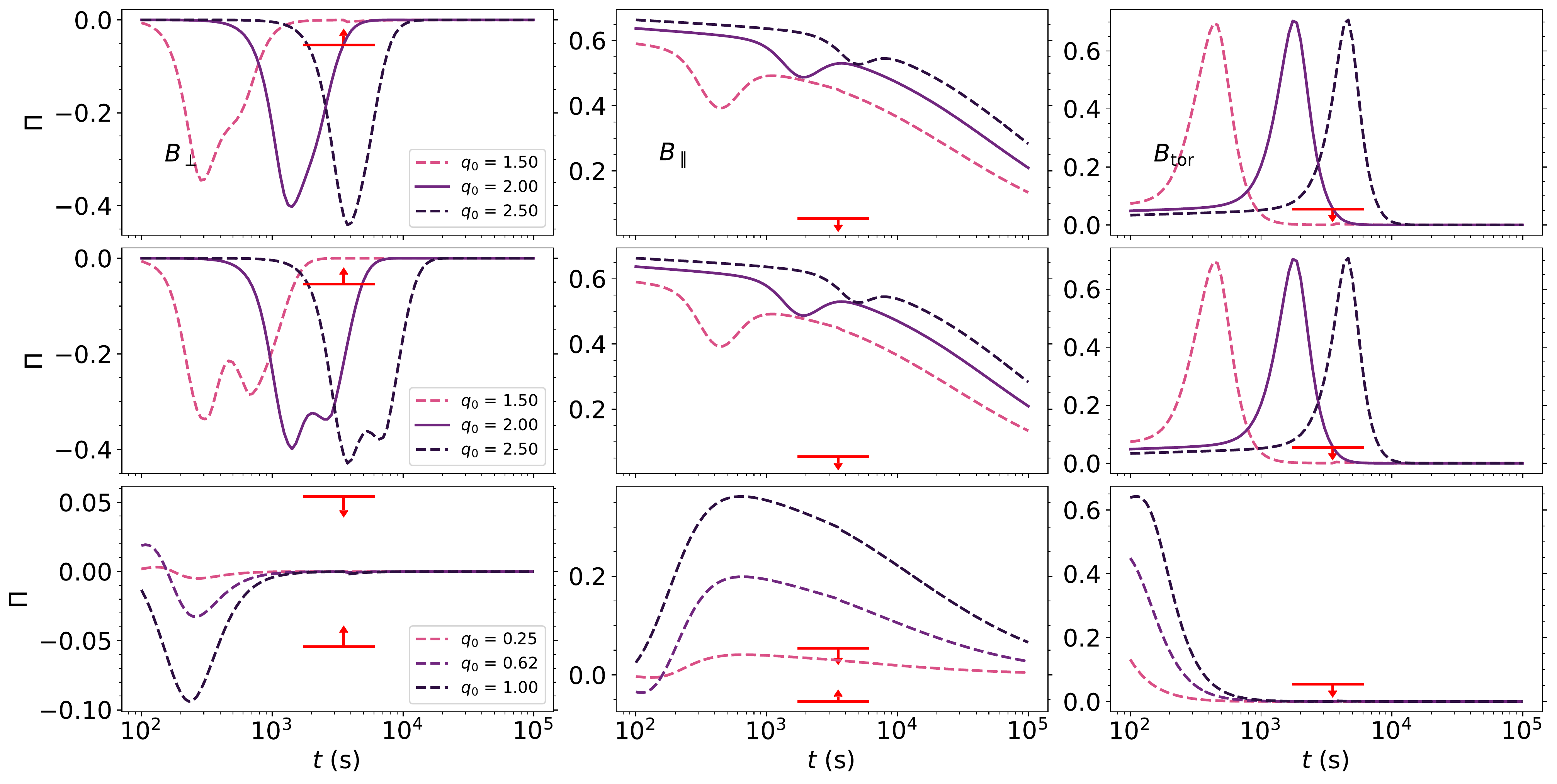}}
\caption{Temporal evolution for the polarization of  GRB 190829A, obtained with the parameters used by \protect\cite{2021MNRAS.504.5647S} (for the narrow jet component) for three configurations of magnetic field  - Perpendicular ($B_\perp$) and Parallel ($B_\parallel$) and Toroidal ($B_{\rm tor}$). Dashed lines represent different values of $q_0$, used for illustration, while solid lines represent the value of $q_0$ obtained from the authors' $\theta_{\rm obs}$ and $\theta_{\rm j}$. Top and middle rows show a non-expanding jet and an expanding jet viewed off-axis, respectively, while bottom row shows an expanding jet viewed within the jet's beaming cone. Upper limits taken from \protect\cite{2022MNRAS.512.2337D}.}
\label{fig:Sato_190829A}
\end{figure}

\Cref{fig:Sato_190829A,fig:Salafia_190829A,fig:Dichiara_190829A,fig:Fraija_190829A,fig:Zhang_190829A,fig:TZhang_190829A} show the polarization degree for three magnetic field configurations, with each figure representing the parameters used by \cite{2021MNRAS.504.5647S}, 
 \cite{2022arXiv220813987S}, \cite{2022ApJ...931L..19S}, \cite{2022MNRAS.512.2337D}, \cite{2021ApJ...918...12F} and \cite{2021ApJ...917...95Z, 2021ApJ...920...55Z}, respectively. The magnetic field configurations are arranged from left to right: $B_\perp$, $B_\parallel$, and $B_{\rm tor}$; and from top to bottom: a non-expanding jet configuration (i.e., homogeneous jet where $\theta_{\rm j}$ does not evolve) viewed off-axis, an expanding jet configuration\footnote{The choice to include both approaches to the homogeneous jet comes from the nearly identical flux light curves, but different polarizations \citep[e.g. see][]{2004MNRAS.354...86R}} (i.e., homogeneous jet where $\theta_{\rm j}=\theta_{\rm j}(t)$, according to \cite{2000ApJ...543...90H} hydrodynamical equations) also viewed off-axis, and the same expanding jet viewed from within the jet's beaming cone ($\theta_{\rm obs} < \theta_{\rm j}$). For all cases, we present the polarization for a range of initial values of $q_0=\frac{\theta_{\rm obs}}{\theta_{\rm j,0}}$, with solid lines as markers for the observation angle reported/assumed by the mentioned authors.

\cite{2021MNRAS.504.5647S, 2022arXiv220813987S} used a narrow jet FS synchrotron model to describe the early afterglow of GRB 190829A. They reported this model to be satisfactory in explaining the multi-wavelength observations in the range from $8\times10^{2}$ to $2\times10^{4}{\rm s}$. We use the most recent best-fit parameters reported in this work (see the first row of \cref{tab:table_par}), with the additional parameters of $\varepsilon_e=3.5\times10^{-2}$, $\varepsilon_B=6\times10^{-5}$, and $\zeta_e =0.2$ and a maximum flux at time $t_{\rm peak} = 2\times10^{3} {\rm s}$. We present in \Cref{fig:Sato_190829A} the polarization evolution considering these parameters. The solid line represents the values presented by \cite{2022arXiv220813987S}. Regardless of the choice of jet (expanding or not), the polarization values returned from these parameters break the upper limits of $\Pi(t \approx [1700 - 6000]{\rm s}) \lesssim 6\%\ $ imposed by \cite{2022MNRAS.512.2337D}.
For an expanding jet, the set of parameters presented by the authors return the following polarization values $\Pi(B_\perp,t \approx 2000\, {\rm s}) \approx 30\%$, $\Pi(B_\perp,t\approx 4000\, {\rm s}) \approx 22.2\%$ and $\Pi(B_\perp,t\approx 6000 \,{\rm s}) \approx 0\%$. For a toroidal field, the polarization observed is $\Pi(B_{\rm tor},t \approx 2000 {\rm s}) \approx 63\%$, $\Pi(B_{\rm tor},t\approx 4000\, {\rm s}) \approx 2\%$ and $\Pi(B_{\rm tor},t\approx 6000\, {\rm s}) \approx 0\%$.
At first glance, this should enough to exclude the possibility of a viewing angle this high ($\theta_{\rm obs}=2\theta_{\rm j}$). However, it is worth noting that polarization is highly dependent on the magnetic field anisotropy, and we only explore the extremely anisotropic cases ($b=0; b\rightarrow\infty$). A more isotropic field, where $b$ is closer to unity, would severely reduce the polarization degree \citep[e.g., see][for the case of GRB 170817A]{2018MNRAS.478.4128G, 2018ApJ...861L..10C, 2020MNRAS.491.5815G, 2021MNRAS.507.5340T} without any interference on the flux light curves (which does not take into consideration the magnetic field's geometry). Exploring this parameter set with a more isotropic configuration could amend the discrepancy between polarization and flux fitting.

\begin{figure}
{\includegraphics[width=\textwidth]{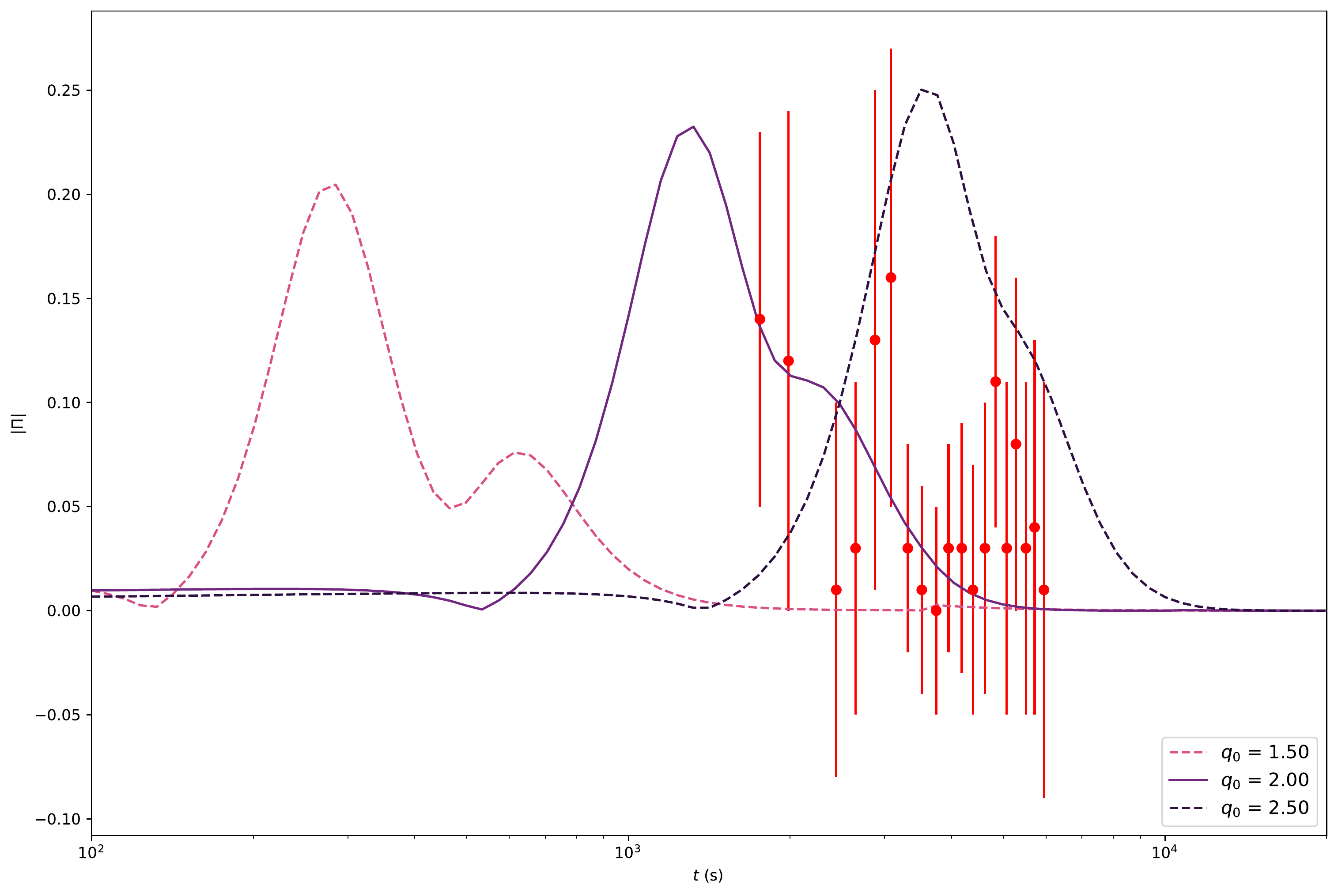}}
\caption{Temporal evolution for the polarization modulus,  of  GRB 190829A, obtained with the parameters used by \protect\cite{2021MNRAS.504.5647S} (for the narrow jet component) for a value of $\eta=0.25$ -- a dominant random component plus a subdominant ordered component. Dashed lines represent different values of $q_0$, used for illustration, while solid lines represent the value of $q_0$ obtained from the authors' $\theta_{\rm obs}$ and $\theta_{\rm j}$. Data points taken from \protect\cite{2022MNRAS.512.2337D}.}
\label{fig:Sato_2}
\end{figure}

\begin{figure}
{\includegraphics[width=\textwidth]{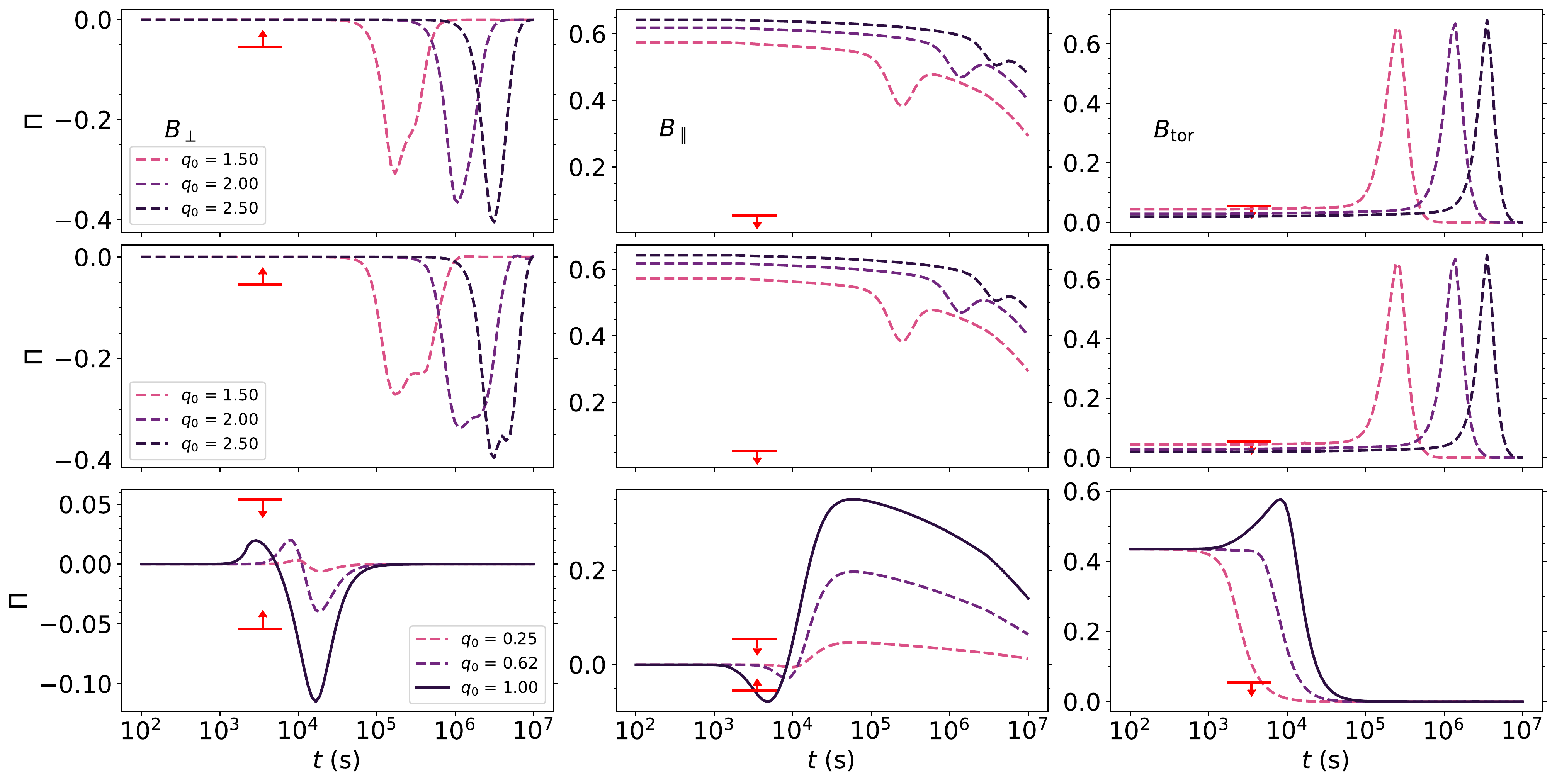}}
\caption{Temporal evolution for the polarization of  GRB 190829A, obtained with the parameters used by \protect\cite{2022ApJ...931L..19S} (for the forward shock component) for three configurations of magnetic field  - Perpendicular ($B_\perp$) and Parallel ($B_\parallel$) and Toroidal ($B_{\rm tor}$). Dashed lines represent different values of $q_0$, used for illustration, while solid lines represent the value of $q_0$ obtained from the authors' $\theta_{\rm obs}$ and $\theta_{\rm j}$. Top and middle rows show a non-expanding jet and an expanding jet viewed off-axis, respectively, while bottom row shows an expanding jet viewed within the jet's beaming cone. Upper limits taken from \protect\cite{2022MNRAS.512.2337D}.}
\label{fig:Salafia_190829A}
\end{figure}

\begin{figure}
{\includegraphics[width=\textwidth]{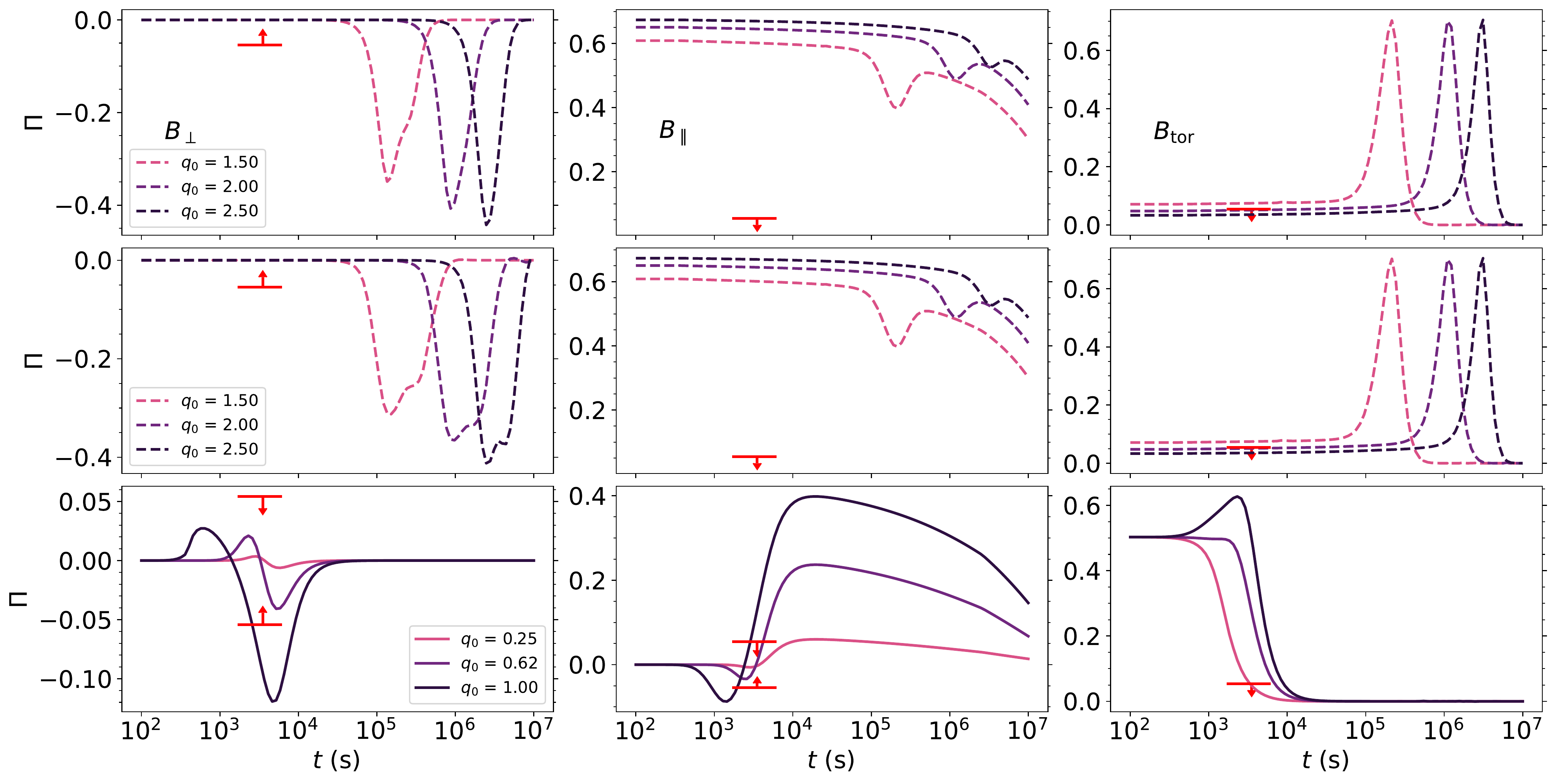}}
\caption{Temporal evolution for the polarization of  GRB 190829A, obtained with the parameters used by \protect\cite{2022MNRAS.512.2337D} (for the forward shock component) for three configurations of magnetic field  - Perpendicular ($B_\perp$) and Parallel ($B_\parallel$) and Toroidal ($B_{\rm tor}$). The values of $\Gamma_0 = 100$ and $n = 5\times10^{-2}$ were assumed. Dashed lines represent different values of $q_0$, used for illustration, while solid lines represent the value of $q_0$ obtained from the authors' $\theta_{\rm obs}$ and $\theta_{\rm j}$. Top and middle rows show a non-expanding jet and an expanding jet viewed off-axis, respectively, while bottom row shows an expanding jet viewed within the jet's beaming cone. Upper limits taken from \protect\cite{2022MNRAS.512.2337D}.}
\label{fig:Dichiara_190829A}
\end{figure}

\begin{figure}
{\includegraphics[width=\textwidth]{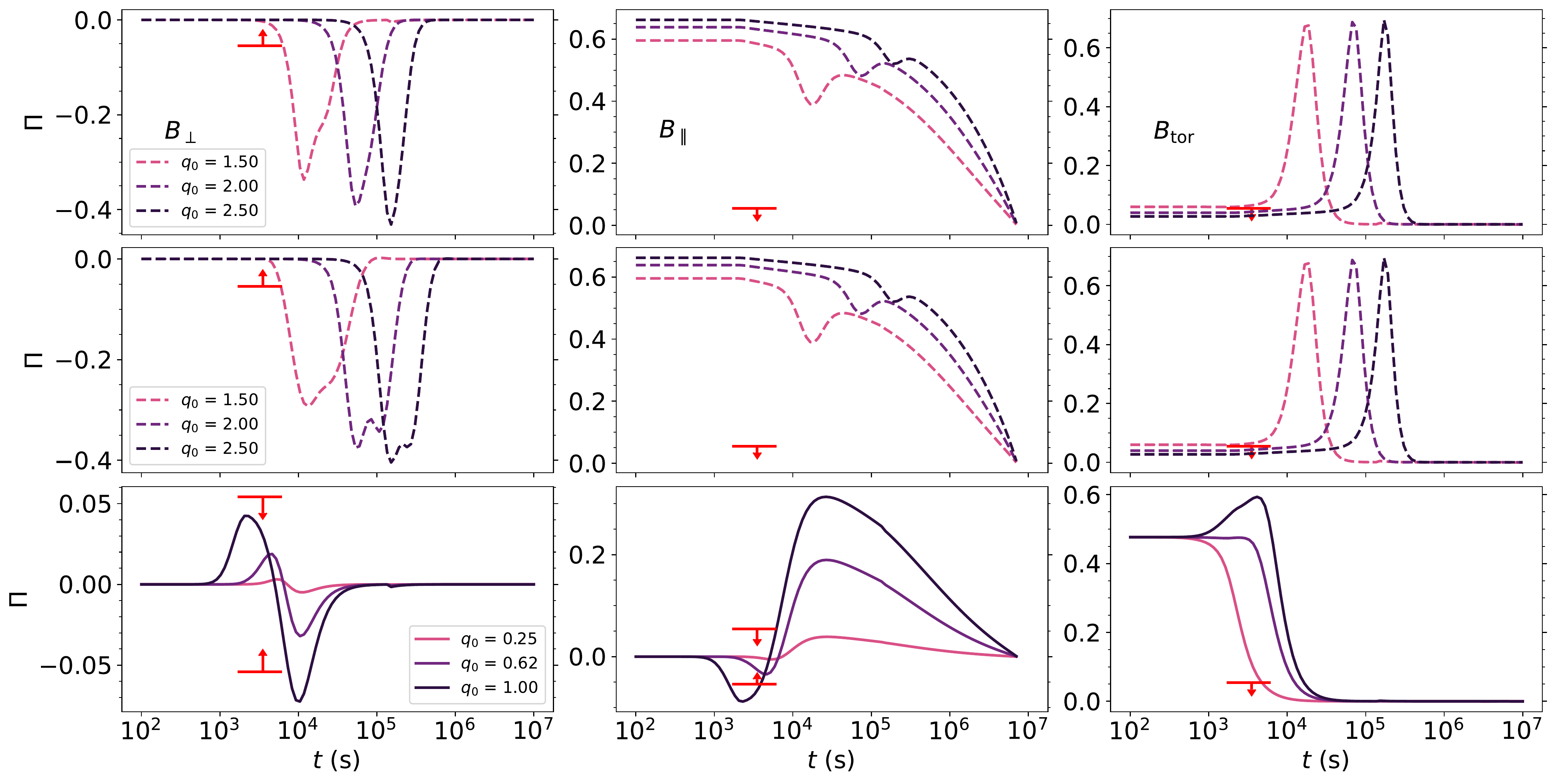}}
\caption{Temporal evolution for the polarization of  GRB 190829A, obtained with the parameters used by \protect\cite{2021ApJ...918...12F} (for the forward shock component) for three configurations of magnetic field  - Perpendicular ($B_\perp$) and Parallel ($B_\parallel$) and Toroidal ($B_{\rm tor}$). Dashed lines represent different values of $q_0$, used for illustration, while solid lines represent the value of $q_0$ obtained from the authors' $\theta_{\rm obs}$ and $\theta_{\rm j}$. Top and middle rows show a non-expanding jet and an expanding jet viewed off-axis, respectively, while bottom row shows an expanding jet viewed within the jet's beaming cone. Upper limits taken from \protect\cite{2022MNRAS.512.2337D}.}
\label{fig:Fraija_190829A}
\end{figure}

\begin{figure}
{\includegraphics[width=\textwidth]{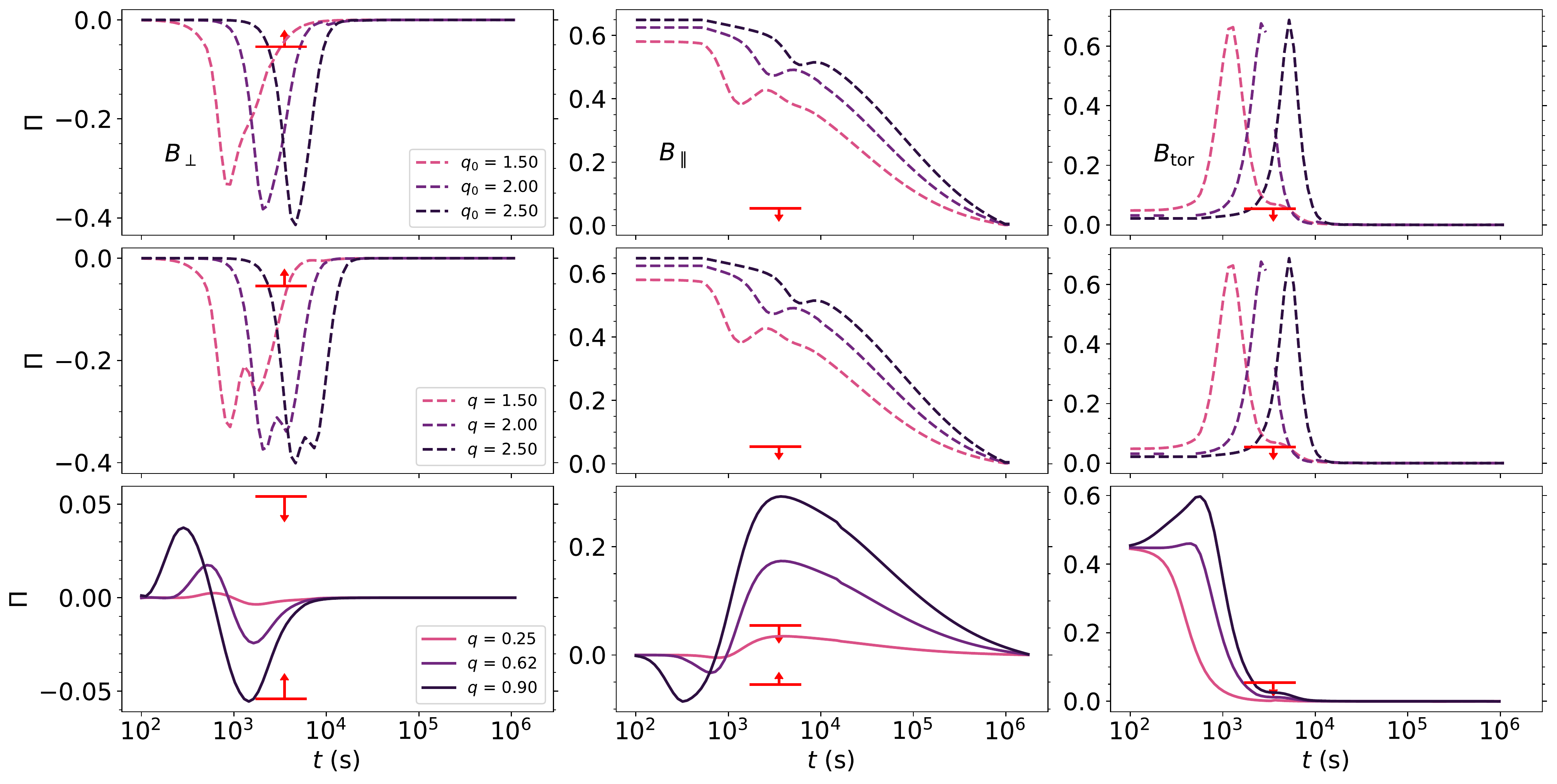}}
\caption{Temporal evolution for the polarization of  GRB 190829A, obtained with the parameters used by \protect\cite{2021ApJ...917...95Z} (for the forward shock component) for three configurations of magnetic field  - Perpendicular ($B_\perp$) and Parallel ($B_\parallel$) and Toroidal ($B_{\rm tor}$). Dashed lines represent different values of $q_0$, used for illustration, while solid lines represent the value of $q_0$ obtained from the authors' $\theta_{\rm obs}$ and $\theta_{\rm j}$. Top and middle rows show a non-expanding jet and an expanding jet viewed off-axis, respectively, while bottom row shows an expanding jet viewed within the jet's beaming cone. Upper limits taken from \protect\cite{2022MNRAS.512.2337D}.}
\label{fig:Zhang_190829A}
\end{figure}

\begin{figure}
{\includegraphics[width=\textwidth]{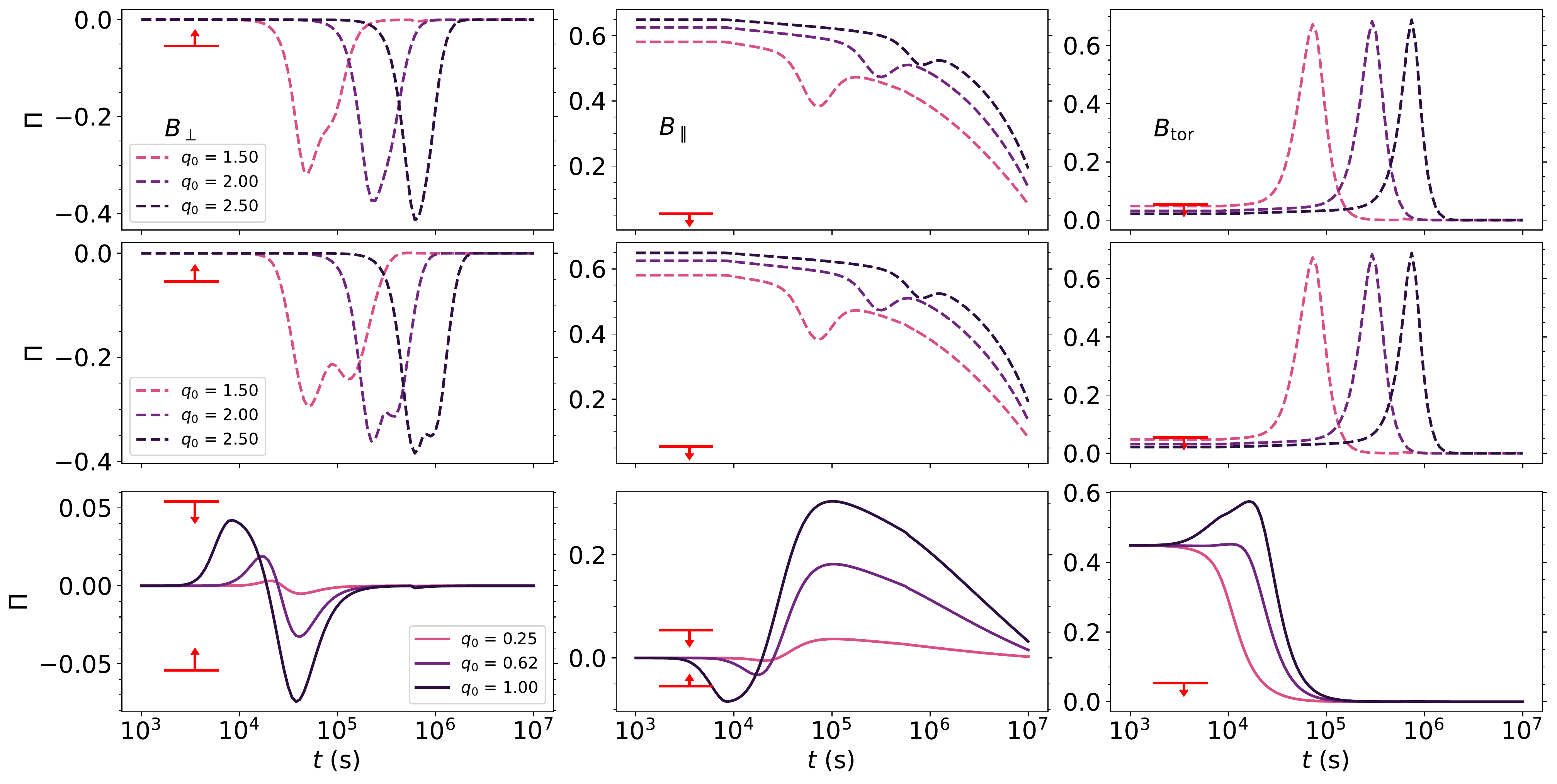}}
\caption{Temporal evolution for the polarization of  GRB 190829A, obtained with the parameters used by \protect\cite{2021ApJ...920...55Z} (for the forward shock component) for three configurations of magnetic field  - Perpendicular ($B_\perp$) and Parallel ($B_\parallel$) and Toroidal ($B_{\rm tor}$). Dashed lines represent different values of $q_0$, used for illustration, while solid lines represent the value of $q_0$ obtained from the authors' $\theta_{\rm obs}$ and $\theta_{\rm j}$. Top and middle rows show a non-expanding jet and an expanding jet viewed off-axis, respectively, while bottom row shows an expanding jet viewed within the jet's beaming cone. Upper limits taken from \protect\cite{2022MNRAS.512.2337D}.}
\label{fig:TZhang_190829A}
\end{figure}

\Cref{fig:Sato_2} shows the polarization calculated for a less extreme anisotropy scenario. To obtain this polarization, we use Eq. 4 of \cite{2003ApJ...594L..83G} to sum the contributions of ordered and random components and take the absolute value:

\begin{eqnarray}
        \Pi = \frac{\eta \Pi_{\rm ord}}{(1+\eta)}\left[1 + \left(\frac{\Pi_{\rm rnd}}{\eta\Pi_{\rm ord}}\right)^2 -\frac{2\Pi_{\rm rnd}}{\eta\Pi_{\rm ord}} \cos{2\delta} \right]^{\frac{1}{2}}.
\end{eqnarray}

This equation introduces two new parameters, $\eta$ and $\delta$, where $\eta = \frac{\expval{B_{\rm ord}^2}}{\expval{B_{\rm rnd}^2}}$ and $\delta$ is the angle between the ordered magnetic field and the jet axis (see Fig. 1 of the aforementioned paper). Information regarding these parameters cannot be obtained from \cite{2021MNRAS.504.5647S, 2022arXiv220813987S}. We explore a possible case where $\eta = 0.25$ and $\delta = \pi/2$. The change to a less extreme anisotropic scenario reduces the polarization significantly and, as is seen in \Cref{fig:Sato_2}, allows to describe the polarization data observed by \cite{2022MNRAS.512.2337D}. This implies the requirement of a sub-dominant ordered component a quarter as strong as the random one, perpendicular to the jet axis, to validate this parameter set. 

\cite{2022ApJ...931L..19S} described the afterglow of GRB 190829A with a dual component; a RS and FS emission. They proposed that the RS emission describes the early multiwavelength afterglow, while the FS emission dominates after $t>1 {\rm day}$. The parameters reported by the authors for the FS emission are found in the second row of \cref{tab:table_par}, and additionally, $\varepsilon_e=3.0^{+2.9}_{-1.7}\times10^{-2}$, $\varepsilon_B=2.5^{+3.5}_{-1.3}\times10^{-5}$, and $\zeta_e =0.04$. The authors also constrained $\Delta\theta<2\,{\rm deg}$ based on a compactness argument and assumed an on-axis (i.e., the observation angle within the beaming cone of the jet) observation angle. For purposes of the calculation we take $\theta_{\rm obs} = \theta_{\rm j}$ as the authors' choice, as indicated by Fig. 5 of their work, but also presented the values for $\theta_{\rm obs} = 0.25\theta_{\rm j}$ and $\theta_{\rm obs} = 0.62\theta_{\rm j}$. \Cref{fig:Salafia_190829A} shows that the authors' chosen values indicate a good fit for the polarization upper limits of $\Pi \lesssim 6\%$, with all values of on-axis $q_0$ returning polarization within the upper limits for the random magnetic field configuration $B_\perp$. We find a polarization degree of $\Pi(B_\perp,t\approx 2000\, {\rm s}) \approx 1.0\%$, $\Pi (B_\perp, t\approx4000\,{\rm s}) \approx 1.2\%$, and $\Pi (B_\perp, t\approx6000\,{\rm s}) \approx -1.4\%$, with $q_0\lesssim 0.62$ required to allow for a polarization originating from a $B_\parallel$ field and $B_{\rm tor}$ ruled out for $q_0 \gtrsim 0.25$. While we limited ourselves to the forward shock synchrotron calculations in this dual component model, the polarization for iC has been explored before \citep{2020MNRAS.491.5815G, 2004MNRAS.347L...1L, 2009ApJ...698.1042T}. These authors have found that polarization for iC is remarkably similar to the corresponding curves for synchrotron emission, with the caveat of a higher possible maximum polarization ($\Pi_{\rm max, syn}\simeq 75\%,\, \Pi_{\rm max, iC} \rightarrow 100\%$); however, the maximum polarization is directly linked to the spectral index $\alpha$ through $\Pi_{max} = (\alpha+1)/(\alpha+1.66)$. As such, with a known $\alpha$ (or $p$) and maximum polarization, the theoretical limit of maximum polarization is of little relevance. The polarization curves would be similar to the ones shown here, with the parameters used to calculate the RS emission.

\cite{2022MNRAS.512.2337D} have also described the afterglow of GRB 190829A with dual component RS plus FS emission. The authors proposed that the rebrightening in the X-ray and optical observations is consistent with the RS component of the emission. The parameters reported by the authors are presented in the third row of \cref{tab:table_par}. Additionally, the authors expected $\varepsilon_e \approx 10^{-1}$ and $\varepsilon_B \approx 10^{-4}$. No fundamental assumptions were made on the observation angle for the emission. However, the authors used the polarization model presented by \cite{2004MNRAS.354...86R} to rule out a homogeneous jet viewed off-axis with a $90\%$ confidence. We then present the curves for $q_0\leq 1$ as the authors' considered values and assume a value of $\Gamma_0=100$, between a minimal value for a typical off-axis jet and the typical values of $\Gamma_0$ for on-axis jets.\footnote{This value is also within the constrained values of \cite{2022ApJ...931L..19S}, based on their compactness argument.} The polarization results are presented in \Cref{fig:Dichiara_190829A}. For all configurations, $q_0\geq1$ is ruled out,  while a value of $q_0= 0.62$ returns the following polarizations $\Pi(B_\perp,t\approx 2000 {\rm s}) \approx 1.8\%$, $\Pi (B_\perp, t\approx4000{\rm s}) \approx -2.5\%$, and $\Pi (B_\perp, t\approx6000{\rm s}) \approx -4.0\%$. For a parallel field configuration, we note that the upper limits rule out $q_0= 1.0, 0.62$, but a value of $q_0<0.62$ can still fit the polarization upper limits. This is in line with the conclusions reached by the authors, who expect that the low radiative efficiency ($<1\%$) could be explained by a jet viewed from $\theta_{\rm obs} < \theta_{\rm j}$. The considerations regarding iC done for \cite{2022ApJ...931L..19S} are the same for this model.

\cite{2021ApJ...918...12F} proposed a synchrotron forward shock model originating from a spin-down millisecond magnetar source to describe the X-ray and optical observations of GRB 190829A, but expect an iC component as necessary to describe the high-energy and VHE photons observed. The parameters reported by the authors are found in the fourth row of \cref{tab:table_par}, with the additional parameters of $\varepsilon_e=0.8^{+0.1}_{-0.1}\times10^{-1}$ and $\varepsilon_B=1.1^{+0.1}_{-0.1}\times10^{-4}$. The authors also assume an on-axis emission to fit the light curves, and we present the polarization assuming $q_0\leq 1$ as the canonical choice. \Cref{fig:Fraija_190829A} shows that, like the parameter set of \cite{2022ApJ...931L..19S}, all choices of $q_0\leq 1$ are well poised to fit the set upper limits. The highest value of on-axis $q_0=1$, an observation angle set at the edge of the jet, returns polarization values of $\Pi(B_\perp,t\approx 2000\, {\rm s}) \approx 4.3\%$, $\Pi (B_\perp, t\approx4000\,{\rm s}) \approx 2.1\%$, and $\Pi (B_\perp, t\approx6000\,{\rm s}) \approx -1.9\%$. For a parallel field configuration, we note that the $q_0=1$ is ruled out, but a value of $q_0\lesssim0.62$ can still fit the polarization upper limits. As such, a choice of $\theta_{\rm obs} = \theta_{\rm j}$ would make $B_\perp$ the sole possible configuration (assuming an extremely anisotropic configuration), while  $\theta_{\rm obs} < \theta_{\rm j}$ would only rule out the globally ordered, toroidal field configuration.

\cite{2021ApJ...917...95Z} proposed a paradigm to explain GRB 190829A. With the condition of a quasi-isotropic ejecta, the authors fit the X-ray, radio, and optical afterglow light curves with a forward shock model, where the afterglow emission is attributed to synchrotron and iC radiation.  The parameters reported by the authors are found in the fifth row of \cref{tab:table_par}, with the additional parameters of $\log_{10}\varepsilon_e = -0.49^{+0.46}_{-0.22} $ and $\log_{10}\varepsilon_B = -3.22^{+1.21}_{-0.80}$. The authors assume an on-axis observation of the burst. As such, we take $q_0\lesssim1$ as their canonical choice. We see from \Cref{fig:Zhang_190829A} that for $q_0=1$ we find $\Pi(B_\perp,t\approx 2000\, {\rm s}) \approx -4.1\%$, $\Pi (B_\perp, t\approx4000\,{\rm s}) \approx -1.0\%$, and $\Pi (B_\perp, t\approx6000\,{\rm s}) \approx 0.0\%$. The values of $q_0 = 1.0,\,0.62$ rule out the globally ordered configurations, which require an $\theta_{\rm obs} \lesssim 0.25\theta_{\rm j}$ to fit the upper limits with these parameters.

\cite{2021ApJ...920...55Z} proposed a combination of the external FS and a late prompt emission to describe the afterglow of GRB 190829A. The authors expected an eiC+SSC mechanism to describe the very-high-energy component by the High Energy Stereoscopic System (H.E.S.S.) observations and the typical FS model accounts for the lower-frequency emissions. The parameters reported by the authors are found in the sixth row of \cref{tab:table_par}, with the additional parameters of $\varepsilon_e=0.39$,  $\varepsilon_e=8.7\times10^{-5}$, and $\zeta_e = 0.34$. The authors assumed an on-axis observation of the burst. As such we take $q_0\lesssim1$ as their canonical choice. \Cref{fig:TZhang_190829A} presents the polarization curves. We see that with this on-axis condition in mind, both a random ($B_\perp$) and ordered ($B_\parallel$) return a polarization within the upper limits of $<6\%$. Taking $q_0=1$, we find $\Pi(B_\perp,t\approx 2000 {\rm s}) \approx 0.0\%$, $\Pi (B_\perp, t\approx4000{\rm s}) \approx 0.5\%$, and $\Pi (B_\perp, t\approx6000{\rm s}) \approx 3.4\%$ and $\Pi(B_\parallel,t\approx 2000 {\rm s}) \approx 0.0\%$, $\Pi (B_\parallel, t\approx4000{\rm s}) \approx -1.4\%$, and $\Pi (B_\parallel, t\approx6000{\rm s}) \approx -5.0\%$. A globally ordered toroidal configuration is ruled out.

\section{Summary and Conclusions}\label{sec7}

In this work, we have applied a semi-analytical FS synchrotron model, analogous with a homogeneous jet, to a polarization model and obtained a set of time-dependent polarization curves. This polarization depends on the parameters associated with the evolution of the bulk Lorentz factor and half-opening angle of synchrotron theory. Additionally, the polarization model depends on the geometry of the magnetic field, and we have explored the extremely anisotropic scenarios ($b=0, \, b\rightarrow\infty$) of a random and two ordered field configurations. We have used different parameter sets of previously published works capable of adequately describing the afterglow of GRB 190829A \citep[e.g. see][]{2021MNRAS.504.5647S,2022arXiv220813987S, 2022ApJ...931L..19S, 2022MNRAS.512.2337D, 2021ApJ...918...12F,2021ApJ...917...95Z, 2021ApJ...920...55Z} to fit the upper limits on polarization set by \cite{2022MNRAS.512.2337D}.

The curves obtained favor the scenario where the observation angle is within the jet's beaming cone ($\theta_{\rm obs} \lesssim \theta_{\rm j}$) while disfavoring an off-axis scenario. This result is in agreement with \cite{2022ApJ...931L..19S, 2022MNRAS.512.2337D, 2021ApJ...918...12F,2021ApJ...917...95Z, 2021ApJ...920...55Z} that predicted GRB 190829A to have been seen on-axis. However, the observations provided by \cite{2022MNRAS.512.2337D} cannot entirely exclude the off-axis scenario. This conclusion comes from the polarization's dependency on magnetic field anisotropy. The presence of a subdominant field (i.e., $b\neq 0,\, b\nrightarrow \infty$) would decrease the observed polarization by a factor $>2$ \citep[model dependent; see][]{2018MNRAS.478.4128G, 2018ApJ...861L..10C, 2020MNRAS.491.5815G, 2021MNRAS.507.5340T}, which is significant enough to reconcile the discrepancy between flux and polarization fitting.

While current evidence indicates that GRB 190829A was seen on-axis, more polarization data and deeper scrutiny of the afterglow fitting would be required to solve the degeneracy present between models properly.

\begin{acknowledgments}
We thank Walas Oliveira, Rodolfo Barniol Duran, Tanmoy Laskar, Paz Beniamini and Bing Zhang for useful discussions. AP acknowledges financial support from CONACyT's doctorate fellowships, NF acknowledges financial support from UNAM-DGAPA-PAPIIT through grant IN106521. RLB acknowledges support from CONACyT postdoctoral fellowships and the support from the DGAPA/UNAM IG100820 and IN105921.
\end{acknowledgments}

\bibliographystyle{mnras}
\bibliography{references}  
\newpage

\begin{table}
\centering \renewcommand{\arraystretch}{2.2}\addtolength{\tabcolsep}{2.4pt}
\caption{Table of parameters reported for GRB 190819A}\label{tab:table_par}
\begin{tabular}{l c c c c c c c}
\hline
\hline
{\large References}& {\large $E\, (10^{53}\,{\rm erg})$} & {\large ${\rm n}\,({\rm cm^{-3}})\,$} & {\large ${\Gamma}_{\rm 0}$}& {\large $\theta_{\rm j}\,\,(\rm deg)$} & {\large $\theta_{\rm obs}\,\,(\rm deg)$} & {\large $p$}   \\

\hline \hline
 \cite{2022arXiv220813987S}& $4 $ & $10^{-2} $ &$350$ & $0.86$ & $1.7$ & $2.44$\\
  \cite{2022ApJ...931L..19S}& $2.5^{+1.9}_{-1.3} $ & $2.1^{+3.7}_{-1.0}\times 10^{-1} $  & $57^{+4}_{-5}$ & $15.4^{+1.3}_{-0.9}$ & -- & $2.01$\\
 \cite{2022MNRAS.512.2337D}& $2\times10^{-1} $ & $10^{-2}-10^{-1} $ &$--$ & $16$ & $--$ & $2.5$\\
  \cite{2021ApJ...918...12F}& $2.4^{+0.2}_{-0.2}\times10^{-2} $ & $1^{+0.1}_{-0.1}\times10^{-1} $ &$34$ & $8$ & $--$ & $2.3^{+0.2}_{-0.2}$\\
  \cite{2021ApJ...917...95Z}& $1.02^{+1.73}_{-0.55}\times10^{-2} $ & $2.18^{+5.76}_{-1.74}\times10^{-1} $ &$35.5^{+16.9}_{-19.0}$ & $3.2$ & $--$ & $2.12^{+0.08}_{-0.17}$\\
  \cite{2021ApJ...920...55Z}& $9.8\times10^{-2} $ & $9\times10^{-2} $ &$25$ & $11.46$ & $--$ & $2.1$\\
\hline
\end{tabular}
\\
\centering{The mean values of the distributions were used for calculation of the polarization curves. Additionally to the $\theta_{\rm obs}$ used by the authors, we include the polarization curves for $q = [0.25, 0.62, 1.50, 2.00, 2.50]$ }
\\
\end{table}

\clearpage

\clearpage

\clearpage

\clearpage

\clearpage

\clearpage

\clearpage

\clearpage

\appendix
\section{Synchrotron FS Off-axis Model}

During the deceleration phase before afterglow emission enters in the observer's field of view, the bulk Lorentz factor is given by Eq. \ref{Gamma_dec_off}.   The minimum and cooling electron Lorentz factors are  given by
{\small
\bary\label{eLor_syn_ism1}
\gamma_m&=& 2.8\times10^{5}\,\left(\frac{1+z}{1.078}\right)^{\frac{3}{2}}\zeta_{e}^{-1} n_{0}^{-\frac{1}{2}} \varepsilon_{e,-1}\theta_{j,5}^{-1}\Delta\theta_{8}^{3}E_{51}^{\frac{1}{2}}t_{3}^{-\frac{3}{2}}\cr
\gamma_c&=&2.3\times10^{3}\,\left(\frac{1+z}{1.078}\right)^{-\frac{1}{2}} n_{0}^{-\frac{1}{2}} (1+Y)^{-1}\varepsilon_{B,-3}^{-1}  \theta_{j,5}\Delta\theta_{8}^{-1}E_{51}^{-\frac{1}{2}} t_{3}^{\frac{1}{2}}\,,
\eary
}
respectively, which correspond to a comoving magnetic field given by {\small $B'\propto \,\left(\frac{1+z}{1.078}\right)^{\frac{3}{2}}\varepsilon_{B,-3}^{\frac{1}{2}}\theta_{j,5}^{-1}\Delta\theta_{8}^{3}E_{51}^{\frac{1}{2}} t_{3}^{-\frac{3}{2}}$}.
%
%
%
The synchrotron spectral breaks can be written as
{\small
\bary\label{En_br_syn_off}
\nu_{\rm m}&=& 434.3\ \mathrm{Hz}\left(\frac{1+z}{1.078}\right)^{2}\zeta_{e}^{-2} n_{0}^{-\frac{1}{2}} \varepsilon_{e,-1}^2 \varepsilon_{B,-3}^{\frac{1}{2}}\theta_{j,5}^{-2}\Delta\theta_{8}^{4}E_{51} t_{3}^{-3}\cr
\nu_{\rm c}&=& 7.3\times10^{-3}\ \mathrm{Hz} \left(\frac{1+z}{1.078}\right)^{-2} n_{0}^{-\frac{1}{2}} (1+Y)^{-2}\varepsilon_{B,-3}^{-\frac{3}{2}}\theta_{j,5}^{2}\Delta\theta_{8}^{-4} E_{51}^{-1} t_{3}
\eary
}
respectively.  The synchrotron spectral breaks in the  self-absorption regime are derived from  $\nu'_{\rm a,1}=\nu'_{\rm c}\tau^{\frac35}_{0,m}$,  $\nu'_{\rm a,2}=\nu'_{\rm m}\tau^{\frac{2}{p+4}}_{0,m}$ and $\nu'_{\rm a,3}=\nu'_{\rm m}\tau^{\frac35}_{0,c}$ with the optical depth given by $\tau_{0,i}\simeq\frac{5}{3}\frac{q_en(r)r}{B'\gamma^5_{\rm i}}$, with $r$ the shock radius \citep{1998ApJ...501..772P}. Therefore, the spectral breaks in the self-absorption regime are given by

{\small
\bary\label{SelfAbsorptionCuts_off}
\nu_{\rm a,1}&\simeq& 9.7\times10^{-15}\ \mathrm{Hz}\left(\frac{1+z}{1.078}\right)^{-4}\zeta_{e}^{\frac85} n_{0}^{\frac{8}{5}} \varepsilon_{e,-1}^{-1} \varepsilon_{B,-3}^{\frac{1}{5}}\theta_{j,5}^{\frac{8}{5}} \Delta\theta_{8}^{-8} E_{51}^{-\frac{4}{5}}t_{3}^{3}\cr
\nu_{\rm a,2}&\simeq& 4.7\times10^{-5}\ \mathrm{Hz} \left(\frac{1+z}{1.078}\right)^{-\frac{2(6-p)}{p+4}}\zeta_{e}^{\frac{2(2-p)}{p+4}} n_{0}^{\frac{10-p}{2(p+4)}} \varepsilon_{B,-3}^{\frac{p+2}{2(p+4)}}\varepsilon_{e,-1}^{\frac{2(p-1)}{p+4}} \theta_{j,5}^{\frac{2(2-p)}{p+4}}\Delta\theta_{8}^{\frac{4(p-6)}{p+4}}E_{51}^{\frac{p-2}{p+4}}t_{3}^{\frac{8-3p}{p+4}}\cr 
\nu_{\rm a,3}&\simeq& 3.3\times10^{-7}\ \mathrm{Hz} \left(\frac{1+z}{1.078}\right)^{-2}\zeta_{e}^{\frac35}(1+Y)n_{0}^{\frac{8}{5}} \varepsilon_{B,-3}^{\frac{6}{5}}\theta_{j,5}^{-\frac{2}{5}} \Delta\theta_{8}^{-4} E_{51}^{\frac{1}{5}} t_{3}\,.
\eary
}

The deceleration time scale $t_{\rm dec}$ can be defined using Eq. \ref{Gamma_dec_off} and the maximum flux is
\bary
F_{\rm max} = 4.5\times10^{-11}\ \mathrm{mJy}\left(\frac{1+z}{1.078}\right)^{-4}\zeta_{e} n_{0}^{\frac{5}{2}} \varepsilon_{B,-3}^{\frac{1}{2}}d_{z,27}^{-2}\theta_{j,5}^{2}\Delta\theta_{8}^{-18} E_{51}^{-1}t_{3}^{6}\,.
\eary
The dynamics of the model post the off-axis phase, generalized for a stratified ambient, are explored in further detail in \cite{2022arXiv220502459F}.

\end{document}